\documentclass[a4paper]{llncs}
\usepackage{amssymb}
\usepackage{amsmath}

\usepackage{graphicx}
\usepackage{tabu}
\usepackage{subfigure}
\usepackage{multirow}
\usepackage[title]{appendix}
\usepackage[normalem]{ulem}
\begin{document}

\title{Generative Modeling and Inverse Imaging of Cardiac Transmembrane Potential}
\titlerunning{Generative Modeling and Inverse Imaging}  

\author{Sandesh Ghimire\inst{1}
\and Jwala Dhamala\inst{1} \and Prashnna Kumar Gyawali\inst{1} \and John L Sapp\inst{2}\and Milan Horacek\inst{2}\and Linwei Wang\inst{1}
}

\authorrunning{S.Ghimire et. al.} 
\institute{ Rochester Institute of Technology, Rochester, NY 14623, USA \\
\email{sg9872@rit.edu},
\texttt{www.sandeshgh.com}
 \and Dalhouse University, Halifax, NS, Canada\\
}

\maketitle              

\begin{abstract}
Noninvasive reconstruction of cardiac transmembrane potential (TMP) 
from surface electrocardiograms (ECG) 
involves an ill-posed inverse problem. Model-constrained regularization is powerful 
for incorporating rich physiological knowledge about spatiotemporal TMP dynamics.  
These models are controlled by high-dimensional physical parameters which, 
if fixed, 
can introduce model errors and reduce the accuracy of TMP reconstruction.  
Simultaneous adaptation of these parameters during TMP reconstruction, 
however, 
is difficult 
due to their high dimensionality. 
We introduce a novel 
model-constrained 
inference framework 
that replaces  
conventional physiological models 
with a deep generative model 
trained to generate TMP sequences 
from low-dimensional generative factors. 
Using a 
variational auto-encoder (VAE)  
with long short-term memory (LSTM) networks, 
we train the VAE decoder to learn the conditional likelihood 
of TMP, 
while the encoder 
learns the
prior distribution of generative factors. 
These two components  
allow us to develop an efficient 
algorithm 
to simultaneously infer 
the generative factors and 
TMP signals from ECG data. 
Synthetic and real-data experiments demonstrate that the presented method 
significantly improve the accuracy of  TMP reconstruction 
compared with 
methods 
constrained  by conventional physiological models
or without physiological constraints.

\keywords{Inverse problem, ECG imaging, Sequential variational auto-encoder, Bayesian inference }
\end{abstract}

\section{Introduction}

Noninvasive electrophysiological (EP) imaging 
{involves}
the reconstruction of cardiac electrical activity from 
high-density {body-surface electrocardiograms} (ECGs) {\cite{gulrajani98}}. 
{It solves an}
ill-posed {inverse} problem 
{that deteriorates as the imaging depth increases 
from the epicardium to the endocardium 
\cite{plonsey07}}. 
One type of 
increasingly utilized regularization 
considers knowledge about the well-defined 
physiological process of cardiac electrical propagation. 
This is often realized 
in a model-constrained approach, 
where the optimization or statistical inference 
of cardiac electrical activity 
is constrained by a pre-defined model 
describing local activation/repolarization   
and its spatial propagation \cite{ghodrati06,vandam09,wang10}. 
Earlier models include 
step jump functions \cite{pullan01}, 
logistic functions \cite{vandam09}, 
and 3D curve models \cite{ghodrati06} 
empirically  
parameterized  
to mimic the physiological process. 
Recently, more expressive 
cardiac EP simulation models 
have also been used 
\cite{wang10,he03}.

These model-constrained approaches 
are afflicted with a common challenge:  
they are controlled by 
high-dimensional parameters 
often associated with 
local tissue properties 
and 
the origin of electrical activation   
that are unknown \emph{a priori}. 
The more expressive the model is, 
the more parameters it has. 
To fix these model parameters in optimization/inference, 
as is common in existing approaches \cite{wang10}, 
model errors  may be introduced 
decreasing the accuracy of the estimated electrical activity \cite{wang10}. 
To adapt these model parameters to the observed data, 
as is desired for accurate inference,  
is however difficult 
due to their 
high-dimensionality  
and nonlinear relationship with the 
observed ECG data 
\cite{ghimire17}. 
 
In this paper, we introduce 
a novel model-constrained inference framework 
that replaces the conventional physiological models with a deep generative model that is trained to generate the spatiotemporal dynamics of transmembrane potential (TMP) from 
a low-dimensional set of \emph{generative factors}. 
These generative factors can be viewed as a low-dimensional abstraction of the high-dimensional physical parameters, 
which allows us to efficiently  
adapt the prior physiological knowledge to the observed ECG data 
(through inference of the generative factors)
for an improved reconstruction of TMP dynamics. 

In specific, the presented method 
consists of two novel contributions. 
First, to obtain a generative model 
that is sufficiently expressive to reproduce the 
temporal sequence of 3D spatial TMP distributions, 
we adopt a novel sequence-to-sequence variational auto-encoder (VAE) \cite{bowman15} 
with cascaded long short-term memory (LSTM) networks.   
This VAE is trained  on a large database of simulated TMP dynamics
originating from various myocardial locations and with a wide range of local tissue properties. 
Second, once trained, 
the VAE decoder 
describes the likelihood 
of the TMP conditioned on a low-dimensional set 
of generative factors, 
while the encoder learns the 
posterior distributions of the generative factors 
conditioned on the training data. 
We utilize these two components 
within the Bayesian inference, and  
present a variation of 
the expectation-maximization (EM) algorithm 
to jointly estimate the generative factors 
and transmural TMP signals 
from observed ECG data. 
In a set of synthetic and real-data experiments,  
we demonstrate that the presented method 
is able to improve the accuracy of transmural EP imaging 
in comparison to 
statistical inference either 
constrained  by a conventional physiological model \cite{wang10}
or without physiological constraints. 

\section{Generative Modeling of TMP via Sequential VAE}
{To learn to generate the spatiotemporal TMP sequences,}
we use a sequential variation of VAE \cite{kingma13}  
{based on the use of LSTM networks \cite{bowman15}.}

\begin{figure}[t!]
\centering
\includegraphics[height=5.8cm]{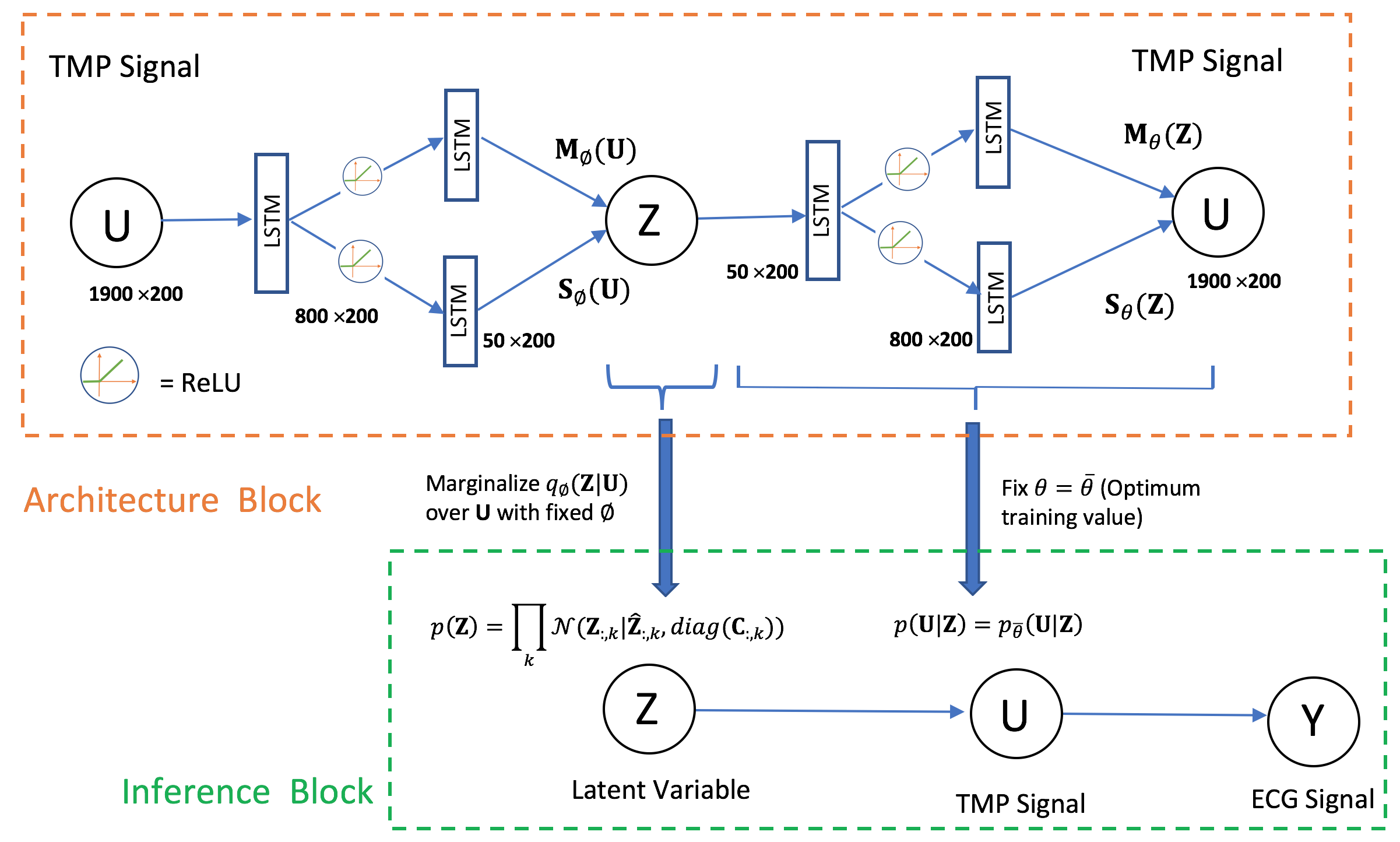}
\caption{\small{Red block: VAE architecture. Green block: graphical model in inference.}}
\label{autoencoder}
\end{figure}

\noindent
\underline{VAE Architecture:}
The architecture of 
the sequential VAE 
is {summarized}
in the red block in 
Fig.~\ref{autoencoder}. 
{Both the encoder and the decoder consists of}
two layers of LSTM, 
where the second layer includes 
separate mean and variance networks. The spatial dimension 
 decreases 
from the original TMP signal $\textbf{U}$ 
to {the} latent representation $\textbf{Z}$, 
{while} 
the temporal relationship is modeled by {the} LSTMs. 
Note that while {the random variables in} a {standard} VAE 
{are vectors}, 
a sequential VAE {deals with} matrices. 
{By defining the} 
conditional distribution of a matrix 
as the product of distributions over its columns, 
{we obtained the likelihood distribution $p_{\theta}(\textbf{U}|\textbf{Z})$ 
and the variational posterior distribution $q_{\phi}(\textbf{Z}|\textbf{U})$} as:
\begin{equation}
p_{\theta}(\textbf{U}|\textbf{Z})=\prod_{k}{\mathcal{N}(\textbf{U}_{:,k}|\textbf{M}_{\theta}(\textbf{Z})_{:,k},diag(\textbf{S}_{\theta}(\textbf{Z})_{:,k}))}
\end{equation}

\begin{equation}
q_{\phi}(\textbf{Z}|\textbf{U})=\prod_{k}{\mathcal{N}(\textbf{Z}_{:,k}|\textbf{M}_{\phi}(\textbf{U})_{:,k},diag(\textbf{S}_{\phi}(\textbf{U})_{:,k}))} 
\end{equation}
where $\textbf{M}_{\phi}(\textbf{U})$ and $\textbf{S}_{\phi}(\textbf{U})$ are 
{output from} 
the mean and variance networks 
of the encoder {parameterized by $\phi$,} 
and $\textbf{M}_{\theta}(\textbf{Z})$ and $\textbf{S}_{\theta}(\textbf{Z})$ are {output from} 
the mean and variance networks of the decoder 
{parameterized by $\theta$}.

\noindent
\underline{{VAE Training:}}
Training of {the VAE} is performed by maximizing the 
{variational} lower bound on the 
{likelihood of the training data given as:}

\begin{equation}
\label{elbo}
\mathcal{L}_{ELB}(\theta,\phi; \textbf{U}^{(i)})=-KL(q_{\phi}(\textbf{Z}|\textbf{U}^{(i)})||p_{\theta}(\textbf{Z}))+E_{q_{\phi}(\textbf{Z}|\textbf{U}^{(i)})}(\log p_{\theta}(\textbf{U}^{(i)}|\textbf{Z})) 
\end{equation}
{where $p_{\theta}(\textbf{Z})$ is an isotropic Gaussian prior. } 
The calculation of {the} KL divergence and cross entropy loss for the presented sequential architecture is carried out in a manner similar to {that described in} \cite{kingma13}. 
The  
training data is generated by  
the Aliev-Panfilov (AP) model \cite{panfilov96}, 
{simulating spatiotemporal TMP sequences originated from}
different 
{ventricular locations}
{with} different tissue properties.

\section{{Transmural EP Imaging}}
The biophysical relationship between {cardiac} TMP, $\mathbf{U}$ and 
{body-surface} ECG, $\mathbf{Y}$ can be described by a 
a linear measurement model: 
$
\textbf{Y}=\textbf{H}\textbf{U}
$, 
where $\mathbf{H}$ is specific to the heart-torso model of an individual. 
To estimate  
$\textbf{U}$ from $\textbf{Y}$ 
is {severely} ill-posed 
{and requires the regularization 
from additional knowledge 
about $\textbf{U}$.}

\subsubsection{Probabilistic Modeling 
{of the Inverse Problem:}}
{We formulate the inverse problem in the form of statistical inference.}
We define the likelihood distribution of $\textbf{Y}$ given $\textbf{U}$ 
{by assuming zero-mean measurement errors with variance $\beta^{-1}$:} 
\begin{equation}
p(\textbf{Y}|\textbf{U}, \beta )=\prod_{k}{\mathcal{N}(\textbf{Y}_{:,k}|\textbf{HU}_{:,k},\beta^{-1}\textbf{I})}
\end{equation}

{To incorporate physiological knowledge about $\textbf{U}$, we 
model its prior distribution conditioned on 
$\textbf{Z}$ 
using the VAE decoder 
with trained parameter $\bar{\theta}$:} 
\begin{equation}
\label{eqn:puz}
p_{\bar{\theta}}(\textbf{U}|\textbf{Z})=\prod_{k}{\mathcal{N}(\textbf{U}_{:,k}|\textbf{M}_{\bar{\theta}}(\textbf{Z})_{:,k},diag(\textbf{S}_{\bar{\theta}}(\textbf{Z})_{:,k}))}
\end{equation}

{To further utilize the knowledge 
about the generative factor $\textbf{Z}$ 
learned by 
the VAE from  
a large training dataset,} 
we also utilize the 
{VAE-encoded} 
marginal posterior distribution of 
$\textbf{Z}$
as its prior distribution 
{in Bayesian inference.} 
{In specific, we} approximate 
samples from
this marginalized distribution to be Gaussian: 
\begin{equation}
\label{eqn:Z}
p(\textbf{Z})=\prod_{k}{\mathcal{N}(\textbf{Z}_{:,k}|\bar{\boldsymbol{Z}}_{:,k},diag(\textbf{C}_{:,k}))}
\end{equation}

With this, we complete the statistical formulation of our problem. 
Our goal is to estimate  
the joint posterior distributions $
p(\textbf{U},\textbf{Z}|\textbf{Y}) \propto
p(\textbf{Y}|\textbf{U})
p(\textbf{U}|\textbf{Z})
p(\textbf{Z}).  
$

\subsubsection{Inference:}
Due to the presence of a deep neural network, the posterior $p(\textbf{U},\textbf{Z}|\textbf{Y})$
is analytically intractable. 
 To address this issue, we note that 
 conditioned on $\textbf{Z}$, 
 the distribution of $\textbf{U}$ is Gaussian in each column; thus,  
 $p(\textbf{U}|\textbf{Y},\textbf{Z})$ 
is analytically available. 
 We leverage this fact and employ 
 a {variant of the} expectation maximization (EM) algorithm to 
{obtain the \emph{maximum a posteriori} (MAP)}
 estimate of $\textbf{Z}$ 
{along with the}
posterior distribution of $\textbf{U}$ 
{given the MAP estimate of $\textbf{Z}$ }. 
 
\noindent
\underline{E-step}: Conditioned on 
{an estimated} value of $\textbf{Z}$ (say $\hat{\textbf{Z}}$), we calculate 
$\small{\hat{p}(\textbf{U}|\textbf{Y},\hat{\textbf{Z}})=}$
$\small{\prod_k \mathcal{N}(\textbf{U}_{:,k}|\hat{\boldsymbol{U}}_{:,k},\hat{\boldsymbol{\Sigma}}_{:,:,k})}$, 
with  
the covariance and mean of the $k^{th}$ column of $\textbf{U}$ as: 
\begin{equation}
\hat{\boldsymbol{\Sigma} }_{:,:,k}=(\beta \textbf{H}^T\textbf{H}+\textbf{D}_k^{-1})^{-1},\hspace{1cm}
\hat{\boldsymbol{U}}_{:,k}=\hat{\boldsymbol{\Sigma}} _{:,:,k}(\beta \textbf{H}^T\textbf{Y}_{:,k}+\textbf{D}_k^{-1}\textbf{m}_k)
\end{equation}
where $\small{\textbf{D}_k=diag(\textbf{S}_{\theta}(\hat{\textbf{Z}})_{:,k})}$, and 
$\small{\textbf{m}_k=\textbf{M}_{\bar{\theta}}(\hat{\textbf{Z}})_{:,k}}$ and $\small{\textbf{S}_{\bar{\theta}}(\hat{\textbf{Z}})_{:,k}}$ are 
the $k^{th}$ column output of the VAE decoder network when $\hat{\textbf{Z}}$ is input to it.

 \noindent
\underline{M-step}: 
Given 
$\small{\hat{p}(\textbf{U}|\textbf{Y},\hat{\textbf{Z}})}$, 
we update $\small{\textbf{Z}}$ by maximizing
$\small {E_{\hat{p}(\textbf{U}|\textbf{Y},\hat{\textbf{Z}})}\log(p(\textbf{Y},\textbf{U},\textbf{Z}))}$
\begin{equation}
\mathcal{L}=E_{\prod_k \mathcal{N}(\textbf{U}_{:,k}|\hat{\boldsymbol{U}}_{:,k},\hat{\boldsymbol{\Sigma}}_{:,:,k})}[\log ( p_{\bar{\theta}}(\textbf{U}|\textbf{Z}))]+\log ( p(\textbf{Z}))+constant 
\end{equation}
Realizing that {a} 
complete optimization of $\mathcal{L}$ with respect to $\textbf{Z}$ would be expensive, we instead take a few gradient descent steps towards the optimum. 
The gradient of the second term 
is analytically available. 
The gradient of the first term is calculated by backpropagation through the decoder network.

The EM steps iterate until convergence, 
at which we obtain both the MAP value of $\textbf{Z}$ and the posterior distribution of $\textbf{U}$ conditioned on $\textbf{Z}$ and $\textbf{Y}$.

\section{Results}
\vspace{-0.25cm}
\subsubsection{Synthetic Experiments:} 

{Synthetic experiments are carried out on 
two image-derived human heart-torso models. 
On each heart, 
the VAE is trained using around 850 simulated TMP signals 
considering approximately 50 different origins of ventricular activation in combination with 17 different tissue property configurations. 
As an initial study, 
here we focus on tissue properties representing 
local regions of myocardial scars 
with varying sizes and locations. 

The presented method 
incorporating the trained VAE model 
is then tested}
on simulated 120-lead ECG data from three {different} settings, 
{each with} 20 experiments. 
{The three settings include}
1) {presence of} myocardial scar not included {in training data,}
2) {origin of ventricular activation}
different from those used in training, 
and 3) both myocardial scar and 
activation origin not seen in training. 
In all experiments, 
the performance of the presented method 
is compared to 0-order Tikhonov regularization {with 
temporal constraint (Greensite method) \cite{greensite98}}
and conventional EP model constrained inference with fixed parameters \cite{wang10}.
  
The reconstruction accuracy is measured with three metrics: 1) normalized RMSE
{given by the ratio of Frobenius norm of the error matrix to that of the truth TMP matrix, 2) 
Euclidean distance between the reconstructed and true origins of ventricular activation, and 3)   
}
Dice coefficient of {the reconstructed $S_1$ and true regions of scar $S_2$ as} =2$|S_1\cap S_2|$/($|S_1| + |S_2|$). 
{In the two physiologically constrained methods}, region of scar is defined based on absence or delay of activation and shortening of action potential duration; in Greensite method, since the reconstructed signal no longer preserves the temporal shape of TMP, the region of scar is defined based on the peak amplitude of the signal.

\textbf{\emph{Computational cost:}}
Training of the VAE takes approximately 40 hours on a 4 GB Nvidia Quadro P1000 GPU. Generation of training data for each heart takes about 7 hours  and inference around 30 minutes on Quadcore CPU.

\begin{figure}[t!]
\centering
\includegraphics[width=.7\linewidth]{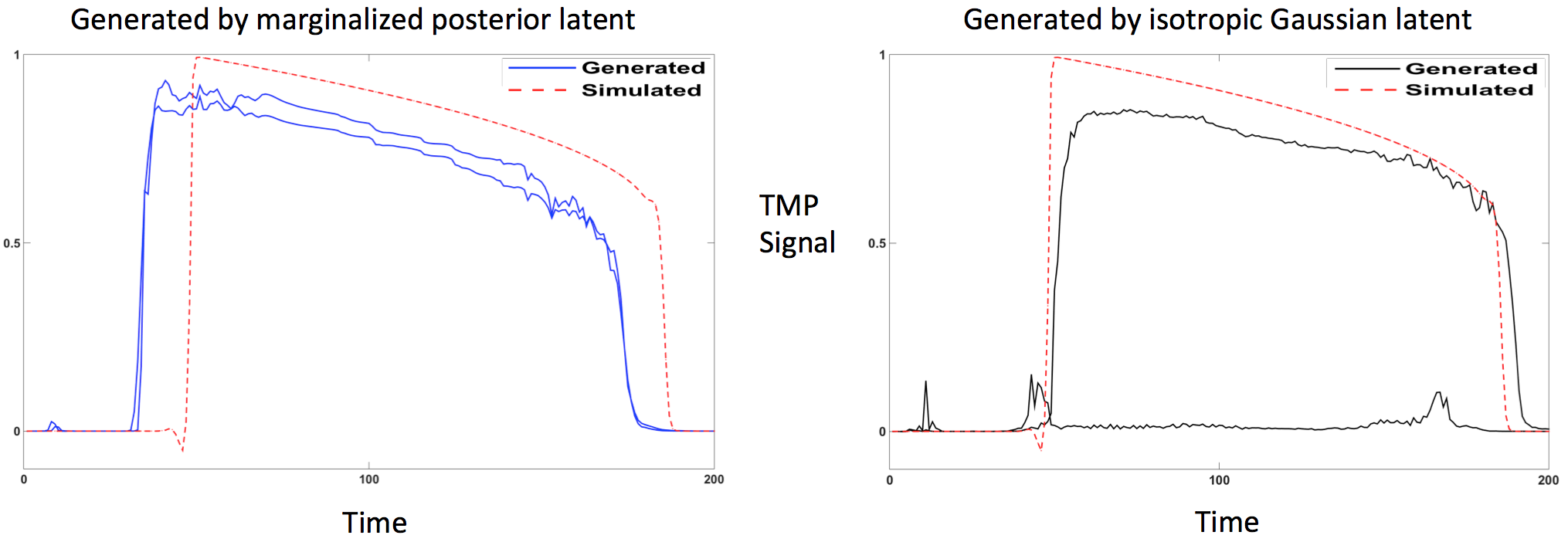}
\caption{Examples of TMP signals generated by 
samples from two different distributions: Left- marginalized posterior density encoded by the VAE ;  Right- isotropic Gaussian.}
\label{sample}
\vspace{-0.4cm}
\end{figure}

\textbf{\emph{TMP generation:}} 
Fig.~\ref{sample} shows examples of 
local TMP signals generated by the trained VAE decoder against TMP signals simulated by the AP model \cite{panfilov96}. 
Note that, 
when generating from a isotropic Gaussian 
(Fig.~\ref{sample} right), 
noisy rather than meaningful TMP signals 
may also be generated. 
In comparison, 
when 
sampling from 
the approximated posterior distribution of $\textbf{Z}$ as described in equation (\ref{eqn:Z}), 
the generated signals 
closely resemble the simulated TMP signals.

\begin{figure}[t!]
\centering
\includegraphics[height=4.2cm]{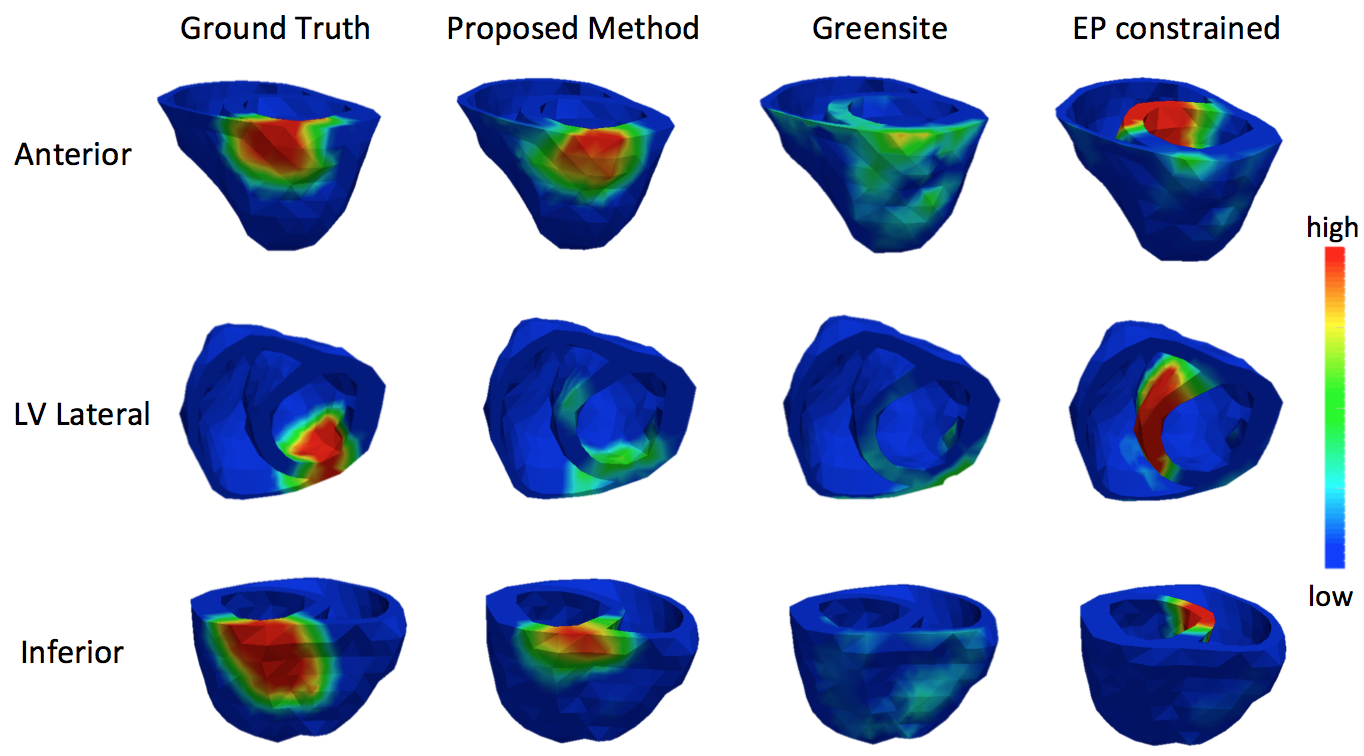}
\caption{Snapshots of early TMP pattern reconstructed by the three 
methods in comparison to the ground truth. The origin of activation  
is noted on the left in each row.}
\label{excitation}
\end{figure}

\begin{figure}[t!]
\centering
\includegraphics[height=4.8cm]{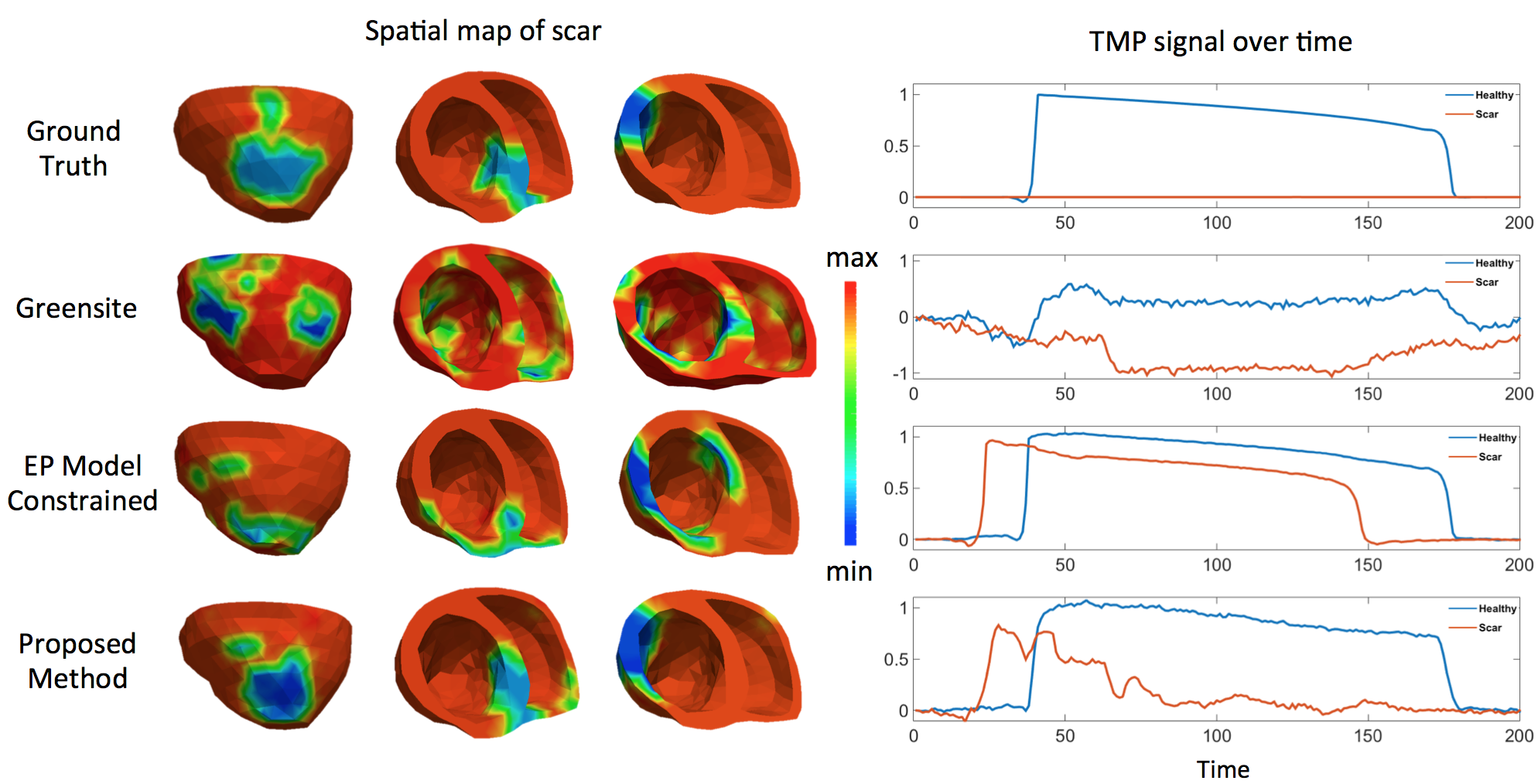}
\caption{Spatial distributions of scar tissues and temporal TMP signals 
obtained by the three methods in comparison to the ground truth. }
\label{tmp}
\vspace{-0.4cm}
\end{figure}

\begin{table}
\centering
\subfigure{
\begin{tabular}{ ||c|c|c|c||} 
\hline
& Greensite & EP constrained & Proposed Method\\
\hline
Normalized RMSE & $1.005 \pm 0.006$ & $0.3 \pm 0.04$ & $\mathbf{0.23 \pm 0.05}$ \\ 
\hline
Dice coefficient& $0.19 \pm 0.04$ & $0.25 \pm 0.09$ & $\mathbf{0.52 \pm 0.2}$ \\ 
\hline
\end{tabular}
}

\subfigure{
\begin{tabular}{ ||c|c|c|c||} 
\hline
& Greensite & EP constrained & Proposed Method\\
\hline
Normalized RMSE & $1.001 \pm 0.003$ & $0.28 \pm 0.05$ & $\mathbf{0.11 \pm 0.08}$ \\ 
\hline
Euclidean Distance & $18.5 \pm 10.96$ & $39.47 \pm 6.3$ & $\mathbf{14.37 \pm 14.0}$ \\ 
\hline
\end{tabular}
}
\subfigure{
\begin{tabular}{ ||c|c|c|c||} 
\hline
& Greensite & EP constrained & Proposed Method\\
\hline
Normalized RMSE & $1.005 \pm 0.003$ & $0.39 \pm 0.03$ & $\mathbf{0.29 \pm 0.09}$ \\ 
\hline
Dice coefficient& $0.20 \pm 0.07$ & $0.21 \pm 0.05$ & $\mathbf{0.48 \pm 0.24} $\\ 
\hline
Euclidean Distance & $18.7 \pm 9.3$ & $65.5 \pm 11.02$ & $\mathbf{17.89 \pm 10.6}$ \\ 
\hline
\end{tabular}
}
\caption{Quantitative accuracy of the three methods in three settings. Test data is simulated with \textbf{1) Top}: scar not in VAE training, \textbf{2) Middle}: activation origin not in training, \textbf{3) Bottom}: both myocardial scar and activation origin not in training.
}
\vspace{-0.2cm}
\end{table}

\textbf{\emph{Imaging TMP from various origins:}}  
Fig.~\ref{excitation} shows a snapshot from 
the early stage of ventricular activation reconstructed 
by the three methods in comparison to the ground truth. 
Since  
the EP model constrained approach 
assumes general sinus-rhythm activation, 
it introduces model error that incorrectly dominates the results. 
 The simple Greensite method, free from erroneous model assumption, 
 actually does a better job in comparison. 
By adapting model generative factors to the data, 
the presented method demonstrates a 
significantly improved ability 
to reconstruct TMP sequence 
resulting from unknown origins. 

\textbf{\emph{Imaging TMP at the presence of myocardial scar:}} 
Fig.~\ref{tmp} shows 
the spatial distribution of scar tissue 
obtained by the three different methods, 
along with temporal TMP signals 
reconstructed in healthy and scar regions, 
in comparison to the ground truth. 
Without prior physiological knowledge, 
the Greensite method 
is not able to preserve the temporal TMP shape, 
resulting in high RMSE error 
as shown in Table 1. 
By thresholding the maximum amplitude of the reconstructed signals, 
the identified region of scar has high false positives 
and resembles poorly with the ground truth. 
The EP model constrained approach 
does a better job in retaining 
the temporal TMP shape. 
However, without prior knowledge about the scar, 
the model error again affects  
the accuracy of TMP reconstruction, 
especially in the early stage of activation 
when a smaller amount of ECG data is available 
for correcting the model error. 
The presented method, 
in comparison, 
is able to recognize the presence of scar tissue, 
adapting the physiological constraint 
for improved TMP reconstructions 
and scar identifications.


%
\textbf{\emph{Summary:}} 
Table 1 summarizes the quantitative comparison of the three methods tested in the three settings as described earlier. 
Although the test cases were not seen by the VAE during training, 
the proposed method shows a significant improvement in inverse reconstruction (paired t-test, p$<$0.001) 
when compared with the other two methods in all settings and metrics except with Euclidean distance using Greensite method, where improvement is only marginal.  
It shows the importance of physiological knowledge and its adaptation to observed data 
during model-constrained inference. 
\vspace{-0.2cm}
\begin{figure}[tb!]
\centering
\includegraphics[height=3.2cm]{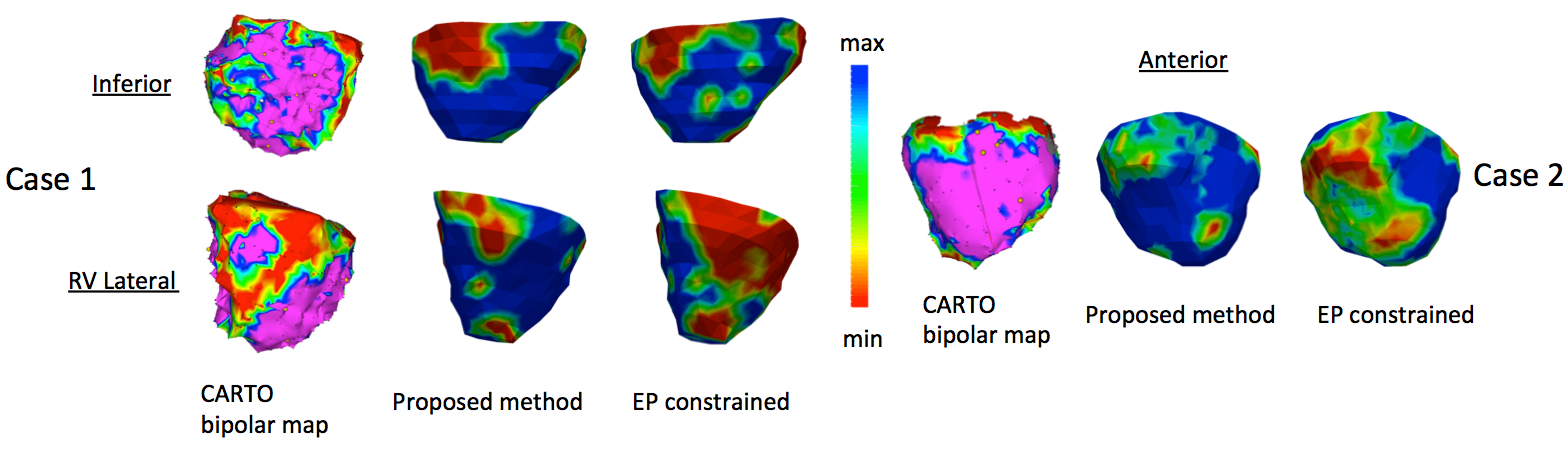}
\caption{Real-data experiments: regions of scar tissues identified by the presented method and conventional EP model constrained method, 
in comparison to bipolar voltage data 
(red: scar core; green: scar border; purple: healthy tissue).}
\label{realdata}
\vspace{-0.3cm}
\end{figure}
 
\subsubsection{Real data Experiments:} 
Two case studies are performed on real-data from patients who underwent catheter ablation due to scar-related ventricular arrhythmia. 
Spatiotemporal TMP is reconstructed from 120-lead ECG data 
using the presented method and the EP model constrained method. 
In Fig.~\ref{realdata}, scar regions (red regions with low voltage) identified from the reconstructed TMP 
are compared with scar regions (red regions) in the \emph{in-vivo} bipolar voltage data. 
In both cases, 
while the scar tissue identified by two methods 
are generally in similar locations, 
the presented method shows less false positives 
and higher qualitative consistency with bipolar voltage maps. 
\vspace{-0.2cm}
\section{Discussion and Conclusions:} 
To our knowledge, 
this is the first work that 
integrates a generative network learned from numerous examples 
into a statistical inference framework 
to allow the adaptation of  
prior physiological knowledge 
via a small number of generative factors. 
The results show the ability of this concept 
to improve model-constrained inference.
Since the present formulation is in a personalized setting, we intend to
extend this architecture to learn a geometry-invariant generative model 
that can be trained on multiple heart models and applied on a new subject. 
\vspace{-0.4cm}
\section*{Acknowledgement}
This work is supported by the National Science Foundation under CAREER
Award ACI-1350374.
\vspace{-0.3cm}
\bibliographystyle{splncs}
\bibliography{bibli1}

\end{document}